\documentclass[sigconf]{acmart}
\pdfoutput=1
\usepackage{graphicx}
\usepackage{bbm} 
\usepackage{xspace}
\usepackage{setspace}
\usepackage{enumitem}
\usepackage{xcolor}
\usepackage{array,booktabs,ragged2e,multirow,bigstrut,diagbox}
\newcolumntype{R}[1]{>{\RaggedLeft\arraybackslash}p{#1}}
\newcolumntype{L}[1]{>{\RaggedRight\arraybackslash}p{#1}}
\newcolumntype{C}[1]{>{\centering\arraybackslash}p{#1}}
\usepackage{pifont}
\newcommand{\vertext}[1]{\rotatebox[origin=c]{90}{#1}}
\usepackage{algorithm}
\usepackage[noend]{algpseudocode}
\makeatletter
\def\BState{\State\hskip-\ALG@thistlm}
\makeatother
\makeatletter
\newcommand{\algmargin}{\the\ALG@thistlm}
\makeatother
\newlength{\whilewidth}
\algdef{SE}[parWHILE]{parWhile}{EndparWhile}[1]
{\parbox[t]{\dimexpr\linewidth-\algmargin}{%
		\hangindent\whilewidth\strut\algorithmicwhile\ #1\ \algorithmicdo\strut}}{\algorithmicend\ \algorithmicwhile}%
\algnewcommand{\parState}[1]{\State%
	\parbox[t]{\dimexpr\linewidth-\algmargin}{\strut #1\strut}}
\newlength{\textfloatsepsave}
\AtBeginDocument{%
  \providecommand\BibTeX{{%
    \normalfont B\kern-0.5em{\scshape i\kern-0.25em b}\kern-0.8em\TeX}}}

\setcopyright{iw3c2w3}

\newcommand{\codeurl}{\url{https://github.com/dhivyaeswaran/hols}\xspace}
\newcommand{\refsec}[1]{Section~\ref{sec:#1}\xspace}
\newcommand{\reffig}[1]{Figure~\ref{fig:#1}\xspace}
\newcommand{\reftab}[1]{Table~\ref{tab:#1}\xspace}
\newcommand{\refeq}[1]{Equation~\eqref{eq:#1}\xspace}
\newcommand{\refalg}[1]{Algorithm~\eqref{alg:#1}\xspace}
\newcommand{\refpr}[1]{Proposition~\ref{pr:#1}\xspace}

\newcommand{\mycomment}[1]{\textit{\color{blue}\small $\triangleright$ #1}}
\newcommand{\mycommentfill}[1]{\textit{\color{blue}\small\hfill $\triangleright$ #1}}

\newcommand{\mywidth}{0.32\linewidth}
\newcommand{\myheight}{1in}

\newcommand{\oof}[1]{\mathcal{O}\left(#1\right)}

\renewcommand{\qed}{\hfill$\blacksquare$}

\newcommand{\ind}[1]{\mathbbm{1}\left[#1\right]\xspace}
\newcommand{\identity}{\mathbf{I}}
\newcommand{\nelem}[1]{|#1|\xspace}
\newcommand{\norm}[1]{||#1||}
\newcommand{\lp}{\textsc{LP}\xspace}
\newcommand{\ls}{\textsc{LS}\xspace}
\newcommand{\hols}{\textsc{HOLS}\xspace}
\newcommand{\holp}{\textsc{HOLS}\xspace}
\newcommand{\gcn}{\textsc{GCN}\xspace}
\newcommand{\nodetovec}{node2vec\xspace}
\newcommand{\tsvm}{\textsc{TSVM}\xspace}
\newcommand{\nvsvm}{node2vec+\tsvm\xspace}

\newcommand{\euemail}{\textsc{EuEmail}\xspace}
\newcommand{\pokec}{\textsc{Pokec}\xspace}
\newcommand{\cora}{\textsc{Cora}\xspace}
\newcommand{\polblogs}{\textsc{PolBlogs}\xspace}
\newcommand{\graph}{\mathcal{G}}
\newcommand{\vertexset}{\mathcal{V}}

\newcommand{\edgeset}{\mathcal{E}}

\newcommand{\edgemat}{\mathbf{W}}

\newcommand{\maxcliquesize}{n}
\newcommand{\nclass}{C}
\newcommand{\cliqueset}{Q}
\newcommand{\cliquemat}[1]{\mathbf{W}^{(#1)}}

\newcommand{\hons}{\mathcal{K}}
\newcommand{\cliquematsmall}[1]{w^{(#1)}}
\newcommand{\bx}{\mathbf{x}}
\newcommand{\by}{\mathbf{y}}
\newcommand{\bX}{\mathbf{X}}
\newcommand{\bY}{\mathbf{Y}}

\newcommand{\loss}{\mathcal{L}}
\newcommand{\edgematsmall}{w}
\newcommand{\lap}{\mathbf{L}}
\newcommand{\symlap}{\tilde{\mathbf{L}}}
\newcommand{\bD}{\mathbf{D}}

\newcommand{\hide}[1]{}

\begin{document}

\title{Higher-Order Label Homogeneity and Spreading in Graphs}

\author{Dhivya Eswaran}
\email{deswaran@cs.cmu.edu}
\affiliation{
	\institution{Carnegie Mellon University}
	\streetaddress{Pittsburgh, PA, USA}
}

\author{Srijan Kumar}
\email{srijan@gatech.edu}
\affiliation{
	\institution{Georgia Institute of Technology}
	\streetaddress{Atlanta, GA, USA}
}

\author{Christos Faloutsos}
\email{christos@cs.cmu.edu}
\affiliation{
	\institution{Carnegie Mellon University}
	\streetaddress{Pittsburgh, PA, USA}
}

\renewcommand{\shortauthors}{Eswaran, Kumar, and Faloutsos}

\begin{abstract}
  Do higher-order network structures aid graph semi-supervised learning? 
Given a graph and a few labeled vertices, labeling the remaining vertices is a high-impact problem with applications in several tasks, such as recommender systems, fraud detection and protein identification. 
However, traditional methods rely on edges for spreading labels, which is limited as all edges are not equal. 
Vertices with stronger connections participate in higher-order structures in graphs, which calls for methods that can leverage these structures in the semi-supervised learning tasks. 

To this end, we propose Higher-Order Label Spreading (\hols) to spread labels using higher-order structures. 
\hols\ has strong theoretical guarantees and reduces to standard label spreading in the base case. 
Via extensive experiments, we show that higher-order label spreading using triangles in addition to edges  is up to 4.7\% better than label spreading using edges alone. Compared to prior traditional and state-of-the-art methods, the proposed method leads to statistically significant accuracy gains in all-but-one cases, while remaining fast and scalable to large graphs. 

\end{abstract}

%
%


\copyrightyear{2020}
\acmYear{2020} 
\acmConference[WWW '20]{Proceedings of The Web Conference 2020}{April 20--24, 2020}{Taipei, Taiwan}
\acmBooktitle{Proceedings of The Web Conference 2020 (WWW '20), April 20--24, 2020, Taipei, Taiwan}
\acmPrice{}
\acmDOI{10.1145/3366423.3379997}
\acmISBN{978-1-4503-7023-3/20/04}

\maketitle

\section{Introduction}
\label{sec:introduction}
\renewcommand{\mywidth}{0.46\columnwidth}
\begin{figure}
	\includegraphics[width=0.9\columnwidth]{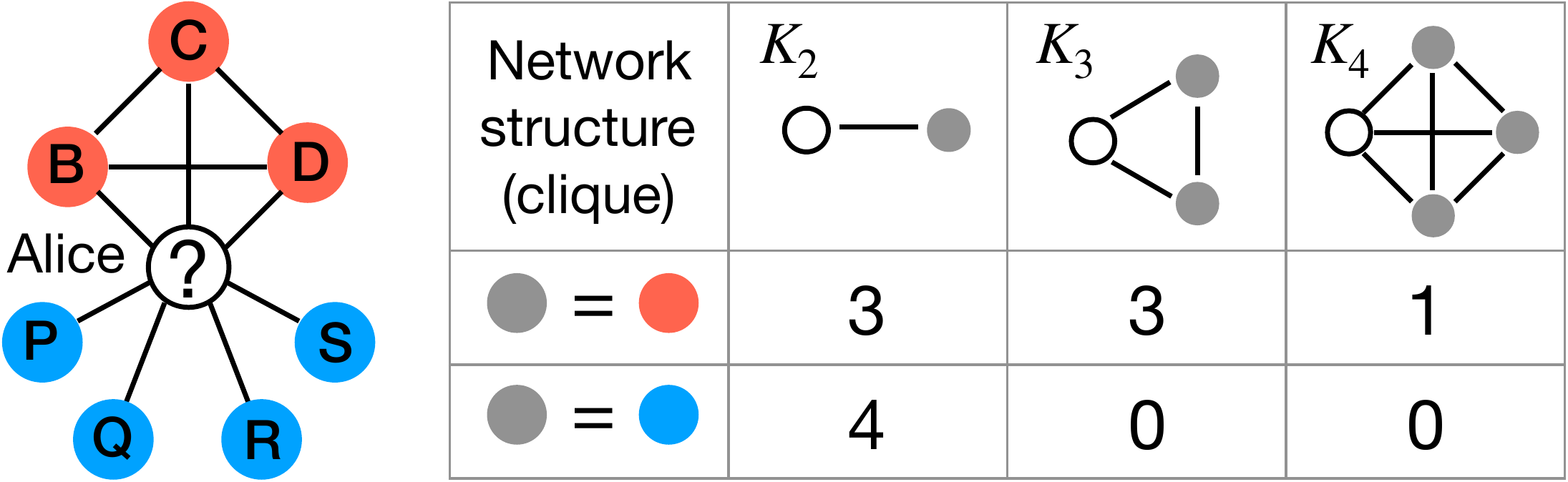}
	\caption{Graph SSL approaches which take only edges into account incorrectly classify the unlabeled central vertex `Alice' as blue. By leveraging higher-order network structures, the proposed \holp correctly labels Alice as red.
	\label{fig:crownjewel}}
\vspace{-2mm}
\end{figure}

Given an undirected unweighted graph and a few labeled vertices, the graph transductive learning or semi-supervised learning (SSL) aims to infer the labels for the remaining unlabeled vertices \cite{DBLP:conf/icml/ZhuGL03,DBLP:conf/nips/ZhouBLWS03,DBLP:conf/pkdd/TalukdarC09,colt04belkin,yedidia2003understanding,iclr17gcn,icml16planetoid,icml19mixhop}.
Graph SSL finds applications in a number of settings: in a social network, we can infer a particular characteristic (e.g. political leaning) of a user based on the information of her friends to produce tailored recommendations; in a user-product bipartite rating network, based on a few manually identified fraudulent user accounts, SSL is useful to spot other fraudulent accounts \cite{DBLP:conf/icwsm/AkogluCF13,DBLP:journals/pvldb/EswaranGFMK17,kumar2018rev2,kumar2019predicting}; SSL can identify protein functions from networks of their physical interaction using just a few labels~\cite{vazquez2003global}.

Traditional graph SSL algorithms leverage a key property of real-world networks: \textit{the homophily of vertices} \cite{albert2002statistical,mcpherson2001birds}, i.e., the nearby vertices in a graph are likely to have the same label.
However, these methods tend to be limited by the fact that all the neighbors of a vertex are not equal.
Consider your own friendship network where you have many acquaintances, but only a few close friends.
In fact, prior research has shown that vertices with a strong connection participate in several higher-order structures, such as dense subgraphs and cliques~\cite{jackson2010social,jackson2012social,sizemore2017classification,hanneman2005introduction}.
Thus, leveraging the higher-order structure between vertices is crucial to accurately label the vertices.

Let us elaborate this using a small friendship network example, shown in Figure~\ref{fig:crownjewel}.
The central vertex, Alice, participates in a closely-knit community with three friends B, C, and D, all of whom know each other. In addition, she has four acquaintances P, Q, R, and S from different walks of her life. Let the vertices be labeled by their ideological beliefs---vertices B, C, and D have the same blue label; and the rest of the vertices have the red label.
Even though Alice has more red connections than blue, the connection between Alice, B, C, and D is stronger as Alice participates in three 3-cliques and one 4-clique with them. In contrast, Alice has no 3- and 4- cliques with P, Q, R, and S.
Owing to the stronger connection with the red nodes, Alice should be labeled red as well.
However, traditional graph SSL techniques that rely on edges alone label Alice as blue~\cite{DBLP:conf/icml/ZhuGL03,DBLP:conf/nips/ZhouBLWS03}.
This calls for graph SSL methods that look beyond edges to leverage the signal present in higher-order structures to label vertices.

Our present work focuses on three key research questions:
\begin{itemize}[leftmargin=*]
	\item \textbf{RQ1.} How do the data reveal that higher-order network structures are homogeneous in labels?
	\item \textbf{RQ2.} How can we leverage higher-order network structures for graph SSL in a principled manner?
	\item \textbf{RQ3.} Do higher-order structures help improve graph SSL?
\end{itemize}

Accordingly, our contributions can be summarized as follows: \\
\textbf{(i)~Analysis:} Through an empirical analysis of four diverse real-world networks, we demonstrate the phenomenon of \textit{higher-order label homogeneity}, i.e., the tendency of vertices participating in a higher-order structure (e.g. triangle) to share the same label.\\
\textbf{(ii)~Algorithm:} We develop Higher-Order Label Spreading (\hols) to leverage higher-order structures during graph semi-supervised learning.
\holp works for any user-inputted higher-order structure and in the base case, is equivalent to edge-based label spreading~\cite{DBLP:conf/nips/ZhouBLWS03}. \\
\textbf{(iii)~Effectiveness:} We show that label spreading via higher-order structures strictly outperforms label spreading via edges by up to 4.7\% statistically significant margin. Notably, \hols\ is competitive with recent deep learning based methods, while running 15$\times$ faster.

For reproducibility, all the code and datasets are available at \codeurl.

\section{Related Work}
\label{sec:relatedwork}
\newcommand{\tick}{\ding{51}}
\begin{table}
	\centering
	\caption{Qualitative comparison of \hols{-}3 (using edges and triangles) with traditional and recent graph SSL approaches\label{tab:salesman}}
	\begin{tabular}{c|cccccc|c}
		\toprule
		\textbf{Desiderata} & \vertext{\lp \cite{DBLP:conf/icml/ZhuGL03}} & \vertext{\ls \cite{DBLP:conf/nips/ZhouBLWS03}} & \vertext{\textsc{BP} \cite{yedidia2003understanding}} & \vertext{Planetoid \cite{icml16planetoid}} & \vertext{\gcn \cite{iclr17gcn}} & \vertext{MixHop \cite{icml19mixhop}} & \vertext{\textbf{\hols{-}3}}\\
		\midrule
		Higher-order structures & & & & \textbf{?} & \textbf{?} & \textbf{?} & \tick\\
		Theoretical guarantees & \tick & \tick & \textbf{?} & & & & \tick\\
		Fast algorithm & \tick & \tick & \tick & & & & \tick\\
		\bottomrule
	\end{tabular}
\end{table}


\textbf{Traditional graph SSL approaches:} By far, the most widely used graph SSL techniques are label propagation \cite{DBLP:conf/icml/ZhuGL03} and label spreading \cite{DBLP:conf/nips/ZhouBLWS03}. Label propagation (\lp) clamps labeled vertices to their provided values and uses a graph Laplacian regularization, while  
label spreading (\ls) uses a squared Euclidean penalty as supervised loss and \textit{normalized} graph Laplacian regularization which is known to be better-behaved and more robust to noise \cite{von2008consistency}.
Both these techniques permit closed-form solution and are extremely fast in practice, scaling well to billion-scale graphs. Consequently, a number of techniques build on top of these approaches, for example, to allow inductive generalization \cite{colt04belkin,DBLP:conf/icml/WestonRC08} and to incorporate certainty \cite{DBLP:conf/pkdd/TalukdarC09}.
When the graph is viewed as pairwise Markov random field, belief propagation (BP) \cite{yedidia2003understanding} may be used to recover the exact marginals on the vertices. BP can handle network effects beyond just homophily; however, it has well-known convergence problems from a practitioner's point of view \cite{DBLP:journals/aim/SenNBGGE08}.
While traditional techniques, in general, show many desirable theoretical properties such as closed-form solution, convergence guarantees, connections to spectral graph theory \cite{DBLP:conf/icml/ZhuGL03} and statistical physics \cite{yedidia2003understanding}, as such, they do not account for higher-order network structures.

\textbf{Recent graph SSL approaches} differ from traditional SSL methods in training embeddings of vertices to jointly predict labels as well as the neighborhood context in the graph. Specifically, Planetoid \cite{icml16planetoid} uses skipgrams, while \gcn \cite{iclr17gcn} uses approximate spectral convolutions to incorporate neighborhood information. MixHop~\cite{icml19mixhop} can learn a general class of neighborhood mixing functions for graph SSL. As such, these do not incorporate \textit{specific higher-order structures} provided by the user. Further, their performance in practice tends to be limited by the availability of `good' vertex features for initializing the optimization procedure.

\textbf{Hybrid approaches for graph SSL:} Another way to tackle the graph SSL problem is a hybrid approach to first extract vertex embeddings using an unsupervised approach such as \nodetovec \cite{node2vec}, DeepWalk \cite{DBLP:conf/kdd/PerozziAS14} or LINE \cite{DBLP:journals/corr/TangQWZYM15} and then use the available labels to learn a transductive classifier such as an SVM \cite{tsvm}. Such methods, however, neither have well-understood theoretical properties nor do they optimize for a single objective in an end-to-end manner.

\textbf{Comparison:} We compare the best performing \hols algorithm (\hols{-}3 which uses triangles in addition to edges) \textit{qualitatively} to prominent SSL approaches in \reftab{salesman} and \textit{quantitatively} via experiments to representative methods from the above categories: \lp and \ls (traditional), \gcn (recent) and \nodetovec + \tsvm (hybrid).

\section{Higher-Order Label Homogeneity}
\label{sec:motivation}
Recent work has shown that graphs from diverse domains have many striking higher-order network structures \cite{benson2016higher} which can be leveraged to improve tasks such as clustering \cite{yin2018higher}, link prediction \cite{DBLP:journals/pnas/BensonASJK18,DBLP:journals/corr/abs-1902-06679} and ranking \cite{DBLP:journals/corr/abs-1906-05059}.
In this section, we motivate the need to consider such structures for semi-supervised learning through empirical analysis of four diverse large real-world networks.
We define and quantify \textit{higher-order label homogeneity}--i.e., the tendency of vertices participating in higher-order structures to share the same label.
We will show that the higher-order label homogeneity is remarkably more common than expected in real-world graphs.

\renewcommand{\myheight}{0.9in}
\renewcommand{\mywidth}{0.05\linewidth}
\newcommand{\mywidthone}{\myheight}
\newcommand{\mywidthtwo}{\myheight}
\newcommand{\mywidththree}{\myheight}
\newcommand{\mywidthfour}{\myheight}
\begin{figure*}[!t]
	\centering
	\resizebox*{\linewidth}{!}{%
		\begin{tabular}{cccc}
			\includegraphics[height=\mywidthone]{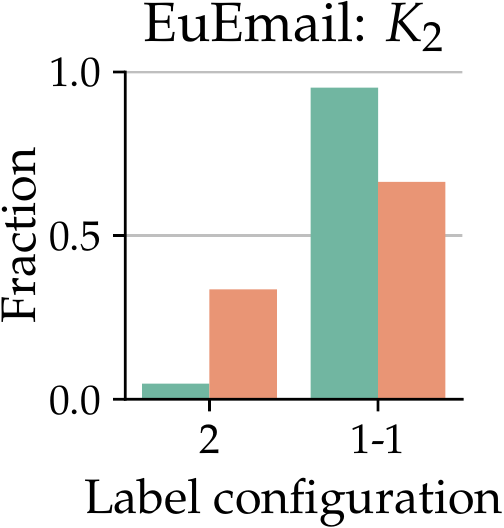} &
			\includegraphics[height=\mywidthtwo]{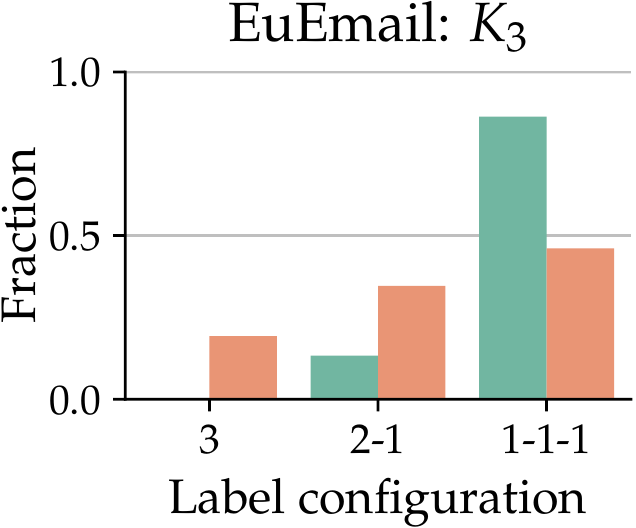} &
			\includegraphics[height=\mywidththree]{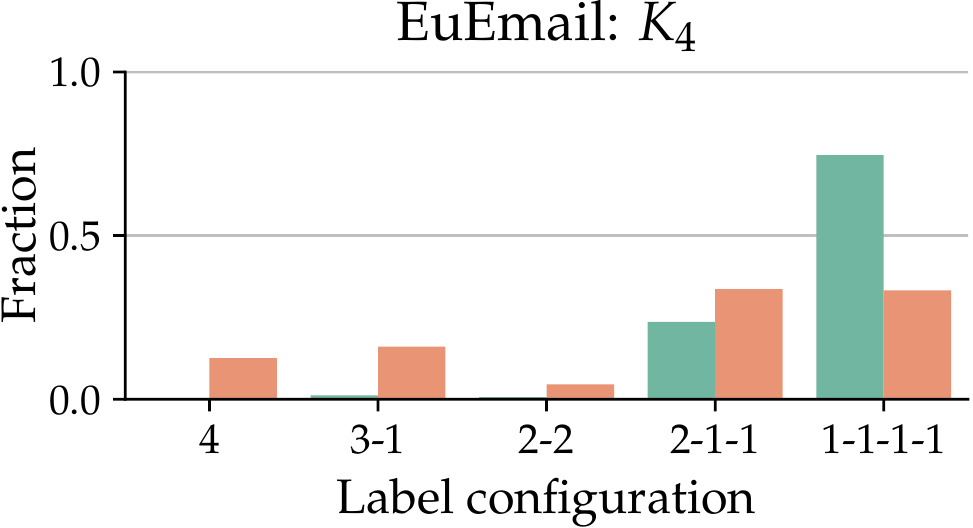} &
			\includegraphics[height=\mywidthfour]{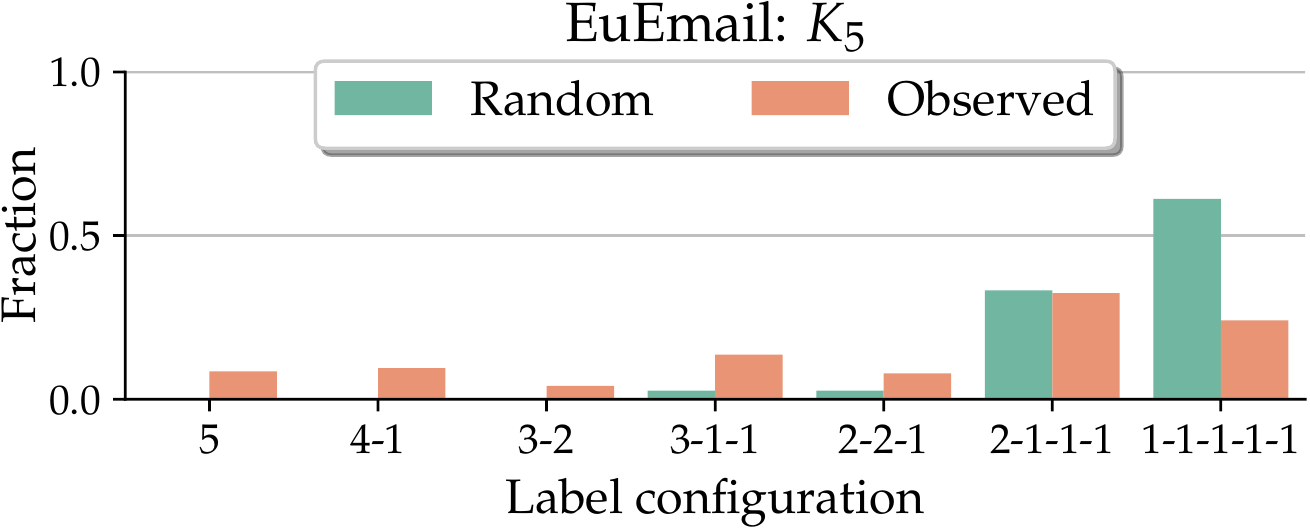} \\
			\includegraphics[height=\mywidthone]{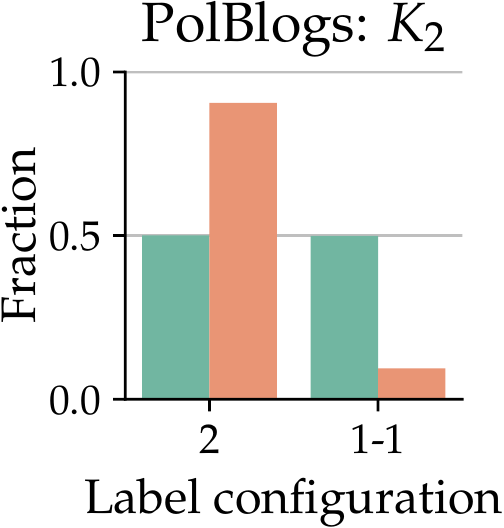} &
			\includegraphics[height=\mywidthtwo]{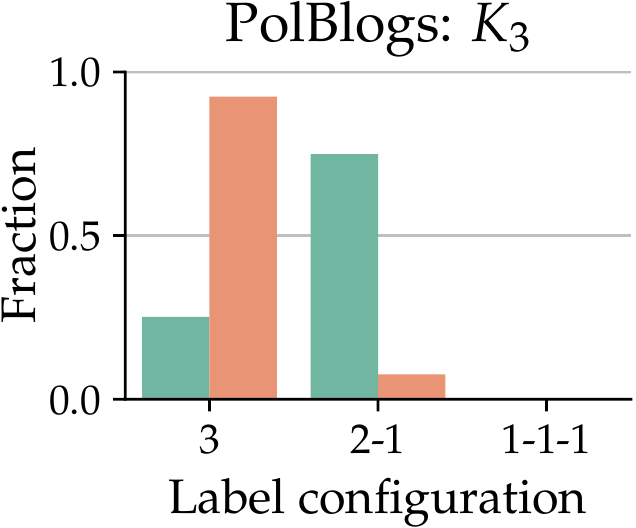} &
			\includegraphics[height=\mywidththree]{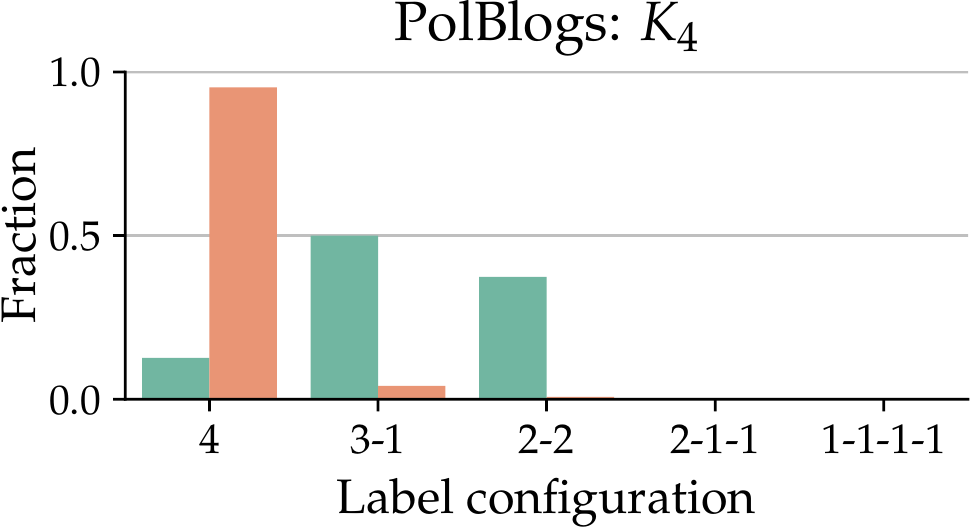} &
			\includegraphics[height=\mywidthfour]{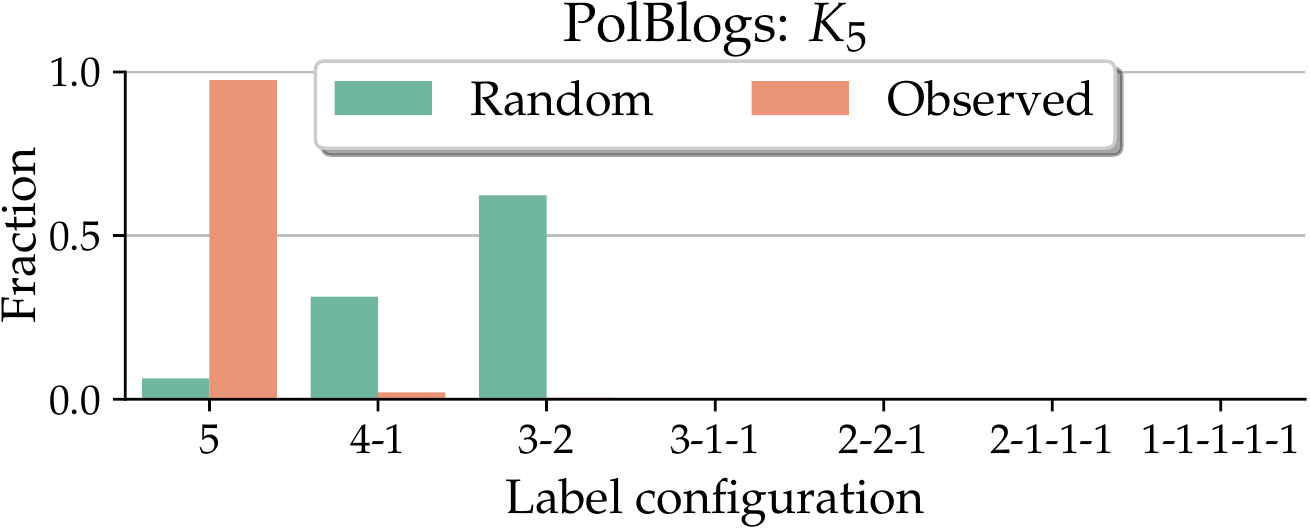}\\
			\includegraphics[height=\mywidthone]{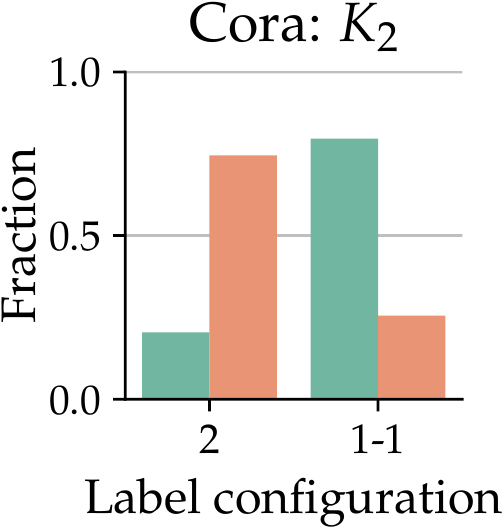} &
			\includegraphics[height=\mywidthtwo]{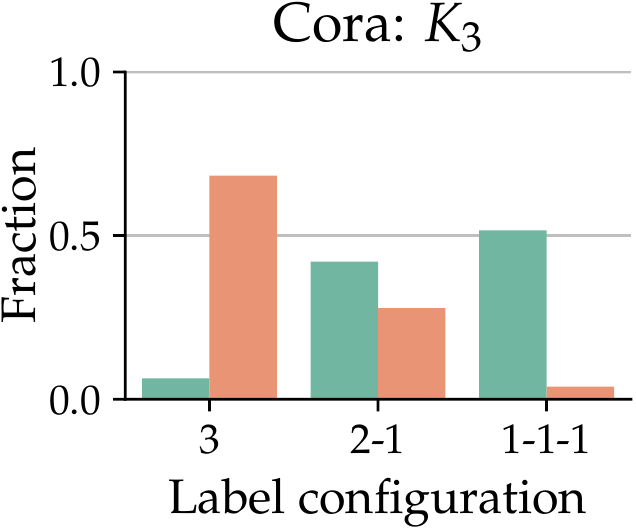} &
			\includegraphics[height=\mywidththree]{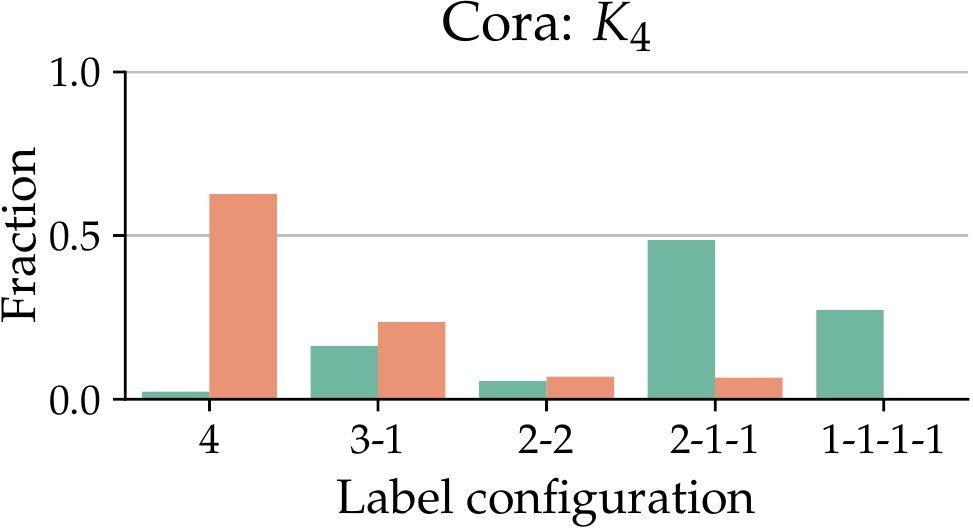} &
			\includegraphics[height=\mywidthfour]{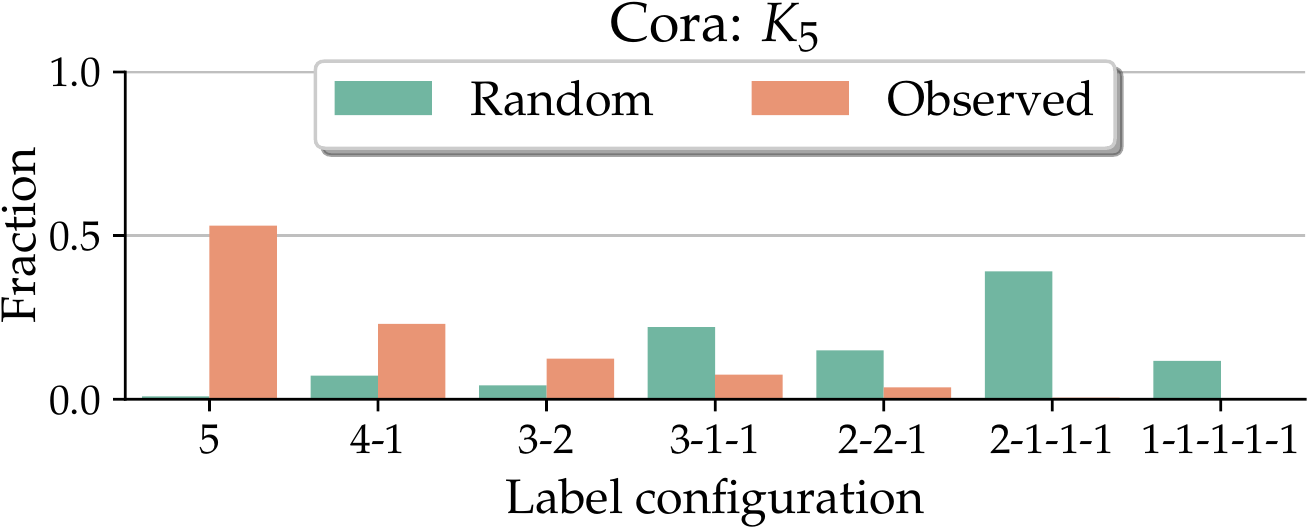}\\
			\includegraphics[height=\mywidthone]{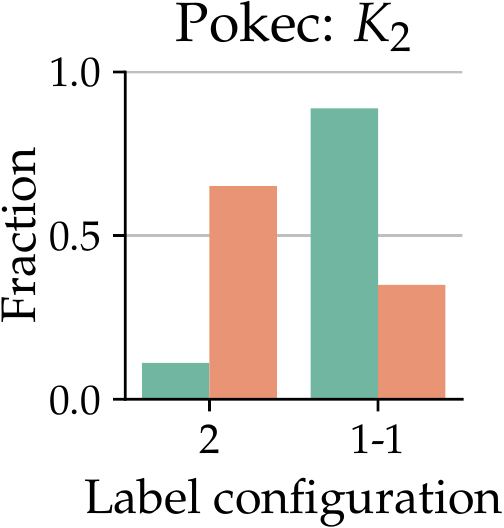} &
			\includegraphics[height=\mywidthtwo]{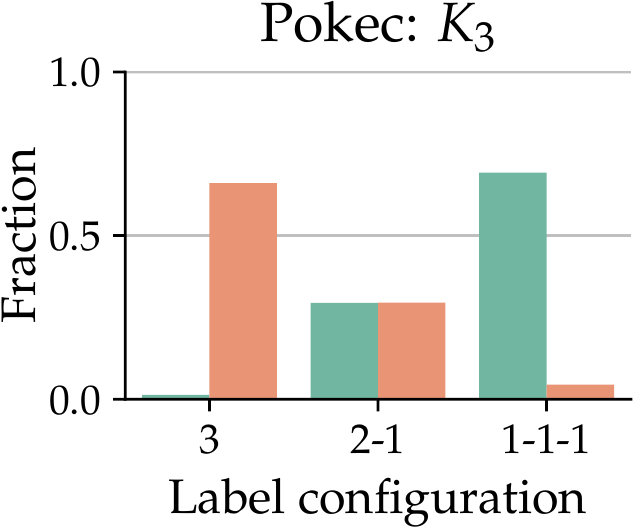} &
			\includegraphics[height=\mywidththree]{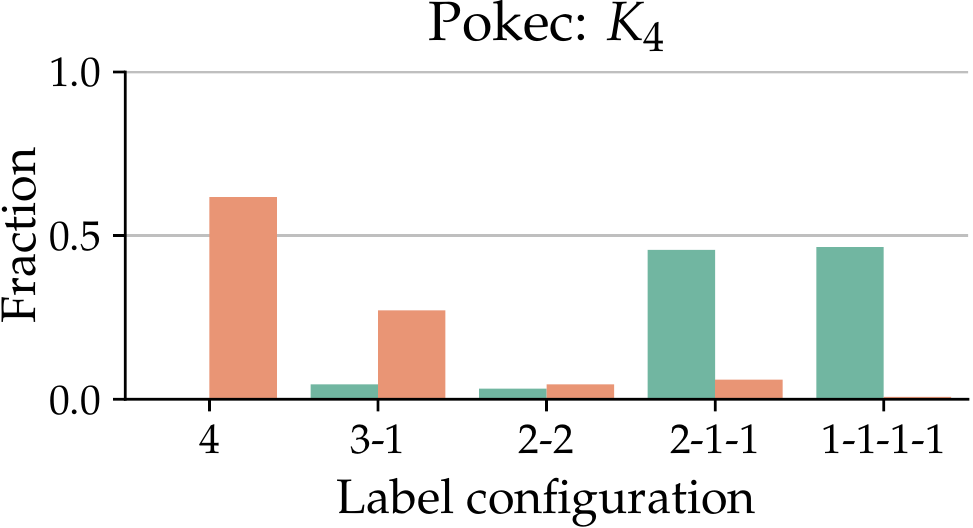} &
			\includegraphics[height=\mywidthfour]{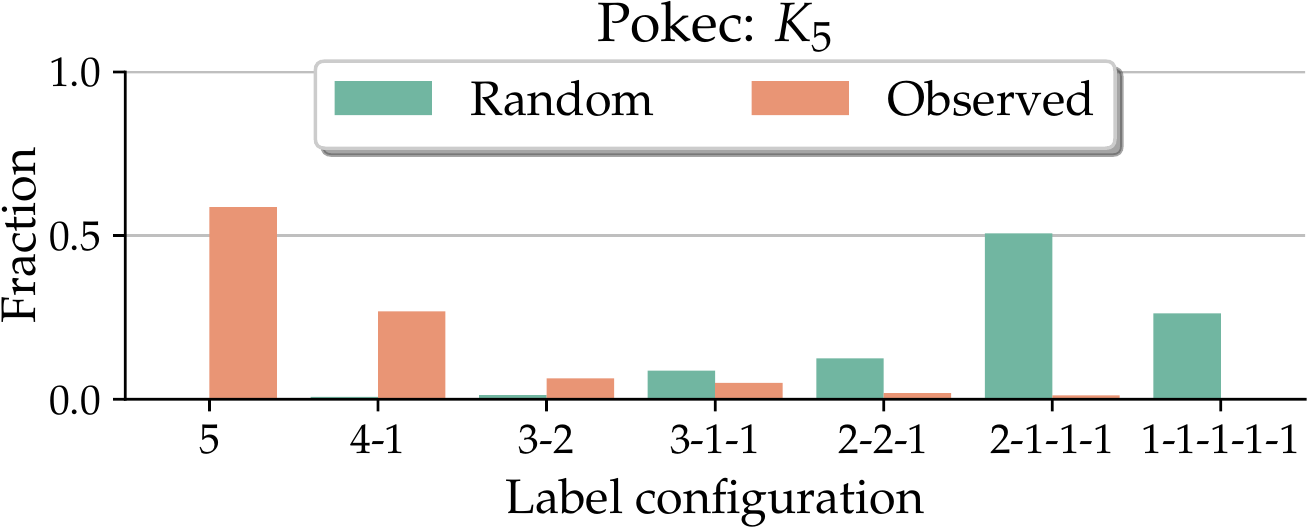}
		\end{tabular}%
	}
	\caption{
		\textit{Prevalence of various label configurations in real-world graphs (orange) relative to a random baseline (green) which shuffles vertex labels fixing the graph structure:}
		We note that more homogeneous label configurations (towards left) are strikingly more prevalent than expected, while less homogeneous label configurations (towards right) are unusually rare.
		\label{fig:labelconfig}}
\end{figure*}

\subsection{Dataset Description}\label{sec:datasets}
\begin{table}[!t]
	\centering
	\caption{Statistics of datasets used\label{tab:datastats}}
	\begin{tabular}{ccccc}
		\toprule
		\textbf{Dataset} & \textbf{Domain} & \textbf{$\nelem{\vertexset}$} & \textbf{$\nelem{\edgeset}$} & \textbf{$\nclass$}\\
		\midrule
		\euemail \cite{DBLP:journals/tkdd/LeskovecKF07} & Email communication & 1005 & 16.0K & 42\\
		\polblogs \cite{DBLP:conf/kdd/AdamicG05} & Blog hyperlinks & 1224 & 16.7K & 2\\
		\cora \cite{DBLP:conf/www/SubeljB13} & Article citations & 23.1K & 89.1K & 10\\
		\pokec \cite{takac2012data} & Friendship & 1.6M & 22.3M & 10\\
		\bottomrule
	\end{tabular}
\vspace{-1mm}
\end{table}
We use four network datasets for our empirical analysis and experiments. \reftab{datastats} summarizes some important dataset statistics.
\begin{itemize}[leftmargin=*]
	\item \textbf{\euemail \cite{DBLP:journals/tkdd/LeskovecKF07}} is an e-mail communication network from a large European research institution. Vertices indicate members of the institution and an edge between a pair of members indicates that they exchanged at least one email. Vertex labels indicate membership to one of the 42 departments.
	\item \textbf{\polblogs \cite{DBLP:conf/kdd/AdamicG05}} is a network of hyperlinks between blogs about US politics during the period preceding the 2004 presidential election. Blogs are labeled as right-leaning or left-leaning.
	\item \textbf{\cora \cite{DBLP:conf/www/SubeljB13}} is a citation network among papers published at computer science conferences. Vertex labels indicate one of 10 areas (e.g. Artificial Intelligence, Databases, Networking) that the paper belongs to based on its venue of publication.
	\item \textbf{\pokec \cite{takac2012data}} is the most popular online social network in Slovakia. Vertices indicate users and edges indicate friendships. From the furnished user profile information, we extract the locality or `kraj' that users belong to and use them as labels.
\end{itemize}

These datasets exhibit homophily \cite{albert2002statistical,mcpherson2001birds}: people typically e-mail others within the same department; blogs tend to link to others having the same political leaning; papers mostly cite those from the same area; people belonging to the same locality are more likely to meet and become friends.
In all cases, we omit self-loops and take the edges as undirected and unweighted.

\subsection{Empirical Evidence}


We now examine the label homogeneity of higher-order $k$-cliques, as they form the essential building blocks of many networks \cite{jackson2010social,jackson2012social,sizemore2017classification,hanneman2005introduction} and moreover, can be counted and enumerated efficiently \cite{DBLP:conf/www/JainS17,DBLP:conf/www/DanischBS18}. We will stick to $k\in\{2,3,4,5\}$ for computational reasons.

\textbf{Methodology.}
We quantify label homogeneity of a given higher-order structure by measuring the distribution over what we term as its \textit{label configurations}.
Simply put, label configuration captures the extent to which participating vertices share the same label and is a function of vertex-label assignments that is invariant under the permutation of vertices and labels. A 2-clique has two label configurations: `2' where both incident vertices have the same label and `1-1' where they have different labels. A 3-clique has three label configurations: `3' where all three vertices have the same label, `2-1' where two of them share the same label and third vertex has a different label and `1-1-1' where each vertex has a different label. Similarly, a 4-clique has 5 label configurations (4, 3-1, 2-2, 2-1-1, 1-1-1-1) and a 5-clique has 7 label configurations (5, 4-1, 3-2, 3-1-1, 2-2-1, 2-1-1-1, 1-1-1-1-1).
Note that not all label configurations may be possible (e.g., 1-1-1 is impossible for a triangle in a 2-class problem) and still fewer may actually occur in practice.

We will now compare the \textit{observed} distribution over label configurations to its commensurate distribution under a \textit{random} baseline or null model, which shuffles vertex labels fixing the graph structure and that marginal distribution of labels.
A priori, there is no reason to expect the \textit{observed} distribution to be any different from \textit{random}. But suppose that the \textit{observed} probability mass for homogeneous label configurations (e.g. `$k$' for $k$-cliques) exceeds that of \textit{random} and vice versa for non-homogeneous label configurations (e.g. `1-$\ldots$-1'); this would suggest higher-order label homogeneity. Similarly, if the opposite occurs, we may conclude that vertices mix \textit{disassortatively} \cite{newman2003mixing} to form higher-order structures.

\textbf{Observations.}
\reffig{labelconfig} plots the observed distribution over $k$-clique label configurations (orange), comparing it to random (green). Under the baseline, most of the probability mass is concentrated on less homogeneous label configurations displayed towards the right, which is expected since vertex labels are assigned at random.

In sharp contrast, the observed distribution is heavily concentrated towards the label configurations on the left. Notably, the most homogeneous label configuration (i.e. `$k$' for $k$-clique, where all participating vertices have the same label), is 1.8-5.9$\times$, 3.7-60$\times$, 7.5-464$\times$, and 15-3416$\times$ more common than expected for $k\in\{2, 3 ,4, 5\}$ respectively. On the other hand, the least homogeneous label configuration (`1-$\ldots$-1', where each participating vertex has a different label) is 1.4-5.3$\times$, 1.8-15$\times$ and 22.2-90$\times$ rarer than expected when possible for $k\in\{2, 3 ,4\}$ respectively. For \cora dataset in particular, the `1-1-1-1-1' label configuration is expected about once in eight or nine 5-clique occurrences ($11.6\%$), but does not occur even once among its over twenty-two thousand 5-cliques.

Overall, our observations establish concrete evidence for the phenomenon of \textit{higher-order label homogeneity}: 
vertices participating in real-world $k$-cliques indeed share the same label to a greater extent than can be explained by random chance.

\section{Higher-Order Label Spreading}
\label{sec:method}
We now derive our higher-order label spreading (\hols) algorithm and show its theoretical properties including the connection to label spreading \cite{DBLP:conf/nips/ZhouBLWS03}, closed-form solution and convergence guarantee.

\subsection{Notation}
\newcommand{\phatvec}{\hat{\mathbf{p}}}
Let $\graph = (\vertexset, \edgeset)$ be a graph with vertex set $\vertexset$ and edge set $\edgeset$. Edges are undirected with $w_{ij}$ representing the edge weight between vertices $i, j \in \vertexset$. Each vertex has a unique label $\ell(i) \in \{1, 2, \ldots, \nclass\}$. 

Let $\hons$ be the set of network structures or motifs (e.g., edges, triangles) which we want to leverage for graph semi-supervised learning.
For a given motif $\kappa \in \hons$, let $\nelem{\kappa}$ denote its \textit{size} which is its number of participating vertices. For example, when $\kappa$ is a triangle, $\nelem{\kappa}=3$. Further, suppose $\cliqueset_\kappa$ is the set of occurrences of a motif $\kappa$ in graph $\graph$ and each such occurrence $q \in \cliqueset_\kappa$ has a weight $w_q$ (e.g. computed as the product of incident edge weights). Use $\ind{\cdot}$ to denote the indicator function, which evaluates to 1 when the enclosed expression is true. For example, $\ind{i \in q}$ is one if the vertex $i$ is part of the subgraph $q$.

\subsection{Generalized Loss Function}\label{sec:alternative}
Let $\by_i \in \{0,1\}^\nclass$ be the one-hot vector of provided label for a labeled vertex $i$ such that $y_{ic}=1$ if vertex $i$ has a label $c$ and is zero otherwise. We propose to infer the final labels $\bx_i\in[0,1]^\nclass$ (where $\sum_c x_{ic} = 1\ \forall\ i\in\vertexset$) by minimizing the following loss function:
\begin{equation}
\loss = (1-\eta) \loss_{s} + \eta \loss_g = (1-\eta)\loss_s + \eta\sum_{\kappa \in \hons} \alpha_\kappa \loss_{g, \kappa} \label{eq:holploss}
\end{equation}
where $\loss_s = \frac{1}{2}\sum_i \norm{\bx_i - \by_i}^2$ is the \textit{supervised loss}, which penalizes the deviation of inferred labels from their provided values, while $\loss_{g, \kappa}$ is the \textit{graph loss} with respect to motif $\kappa$, which enforces the inferred labels to be `smooth' over all occurrences of $\kappa$ as:
\begin{equation}
\loss_{g, \kappa} = \frac{1}{2}\sum_{q\in\cliqueset{\kappa}} w_q \sum_{i,j\in q} \norm{\bx_i - \bx_j}^2
\end{equation}
A parameter $\eta \in (0, 1)$ trades off supervised and graph losses, while $\alpha_\kappa \in (0, 1)$ captures the importance weight of $\kappa$ in semi-supervised learning. Note $\sum_{\kappa\in\hons} \alpha_\kappa = 1$.
Now, define $\kappa$-participation matrix as $\cliquemat{\kappa} = [\cliquematsmall{\kappa}_{ij}]$ where each entry $\cliquematsmall{\kappa}_{ij}$ denotes the total weight of $\kappa$-motifs that vertices $i$ and $j$ participate in. We have:
\begin{equation}
\cliquematsmall{\kappa}_{ij} = \sum_{q\in\cliqueset_\kappa} w_q\cdot \ind{i\in q \land j \in q}
\end{equation}
Observe that each pairwise loss term $\norm{\bx_i-\bx_j}^2$ in \refeq{holploss} appears with a total weight $\edgematsmall'_{ij}$ given by
$\edgematsmall'_{ij} = \sum_{\kappa\in\hons} \alpha_\kappa\cliquematsmall{\kappa}_{ij}$
using which we may simplify the graph loss as:
\begin{equation}
\loss_g = \frac{\eta}{2} \sum_{i, j} \edgematsmall'_{ij} \norm{\bx_i-\bx_j}^2\label{eq:simpleobj}
\end{equation}
Thus, \refeq{simpleobj} establishes that the graph loss from \refeq{holploss} is equivalent to that of label propagation \cite{DBLP:conf/icml/ZhuGL03} on a modified graph with adjacency matrix $\edgemat' = \sum_{\kappa\in\hons}\alpha_\kappa\cliquemat{\kappa}$ where each edge of the original graph has been re-weighted according to the total weight of $\kappa$-motifs it participates in, scaled by its importance $\alpha_\kappa$, and finally summed over all motifs $\kappa \in \hons$ of interest. We will use this connection to derive a closed-form solution to \holp.

{
\small
\begin{algorithm}[!t]
	\caption{\textbf{Higher-Order Label Spreading (\hols)}}\label{alg:hols}
	\begin{algorithmic}[1]
		\Statex \textbf{Input:} graph $\graph=(\vertexset, \edgeset)$, number of classes $\nclass$, set of labeled vertices $\vertexset_l \subset \vertexset$ and their labels $\ell:\vertexset_l\rightarrow\{1, \ldots, \nclass\}$ (at least one labeled vertex per class)
		\Statex \textbf{Parameters:} motif set $\hons$, motif weights $\alpha_\kappa \in (0, 1)$ such that $\sum_{\kappa \in \hons}\alpha_\kappa = 1$, weight $\eta \in (0, 1)$ for supervised loss
		\Statex \textbf{Output:} final label assignments $\ell^\ast(i)$ for all vertices $i \in \vertexset$
		\Procedure{\textsc{HigherOrderLabelSpreading}}{$\graph, \vertexset_l, \ell, \hons, \alpha,\eta$}
		\Statex \mycomment{Construct higher-order normalized graph Laplacian for regularization}
		\For {$\kappa \in \hons$}
		\parState {Construct $\kappa$-participation matrix $\cliquemat{\kappa} = [\cliquematsmall{\kappa}_{ij}]$}
		\Statex \mycommentfill{$\cliquematsmall{\kappa}_{ij}:$ total weight of $\kappa$-motifs where $i$ and $j$ appear together}
		\EndFor
		\State $\edgemat' \leftarrow \sum_{\kappa\in\hons} \alpha_\kappa \cliquemat{\kappa}$ 
		\State $\bD' \leftarrow \textit{diag}(d'_{ii})$ where $d'_{ii} = \sum_{j} \edgematsmall'_{ij}$
		\State $\symlap' \leftarrow \bD'^{-1/2}\edgemat'\bD'^{-1/2}$
		\Statex \mycomment{Construct label matrices $\bY = [y_{ic}]$ (prior) and $\bX = [x_{ic}]$ (inferred)}
		\State $\bY \leftarrow \mathbf{0}_{\nelem{\vertexset}\times\nclass}$\
		\State $y_{i\ell(i)} \leftarrow 1\ \ \forall\ i \in \vertexset_l$
		\State $\bX \leftarrow \bY$
		\Statex \mycomment{Label inference using \hols}
		\While {not converged}
		\State $\bX \leftarrow \eta (\identity-\symlap')\bX + (1-\eta)\bY$ \mycommentfill{\refeq{update}}
		\EndWhile
		\State $\ell^\ast(i) \leftarrow {\arg}{\max}_{c}\ x_{ic} \ \ \forall\ i \in \vertexset$
		\State \textbf{return } $\ell^\ast$
		\EndProcedure
	\end{algorithmic}
\end{algorithm}
}

\subsection{Closed-Form and Iterative Solutions}
Let $\bY = [\by_1 \ldots \by_N]^T$ and $\bX = [\bx_1 \ldots \bx_N]^T$ be the $N\times\nclass$ matrices of prior and inferred labels where $\nelem{\vertexset}=N$ is the total number of vertices. Let $\bD' = [d'_{ij}]$ be the diagonal degree matrix for the modified graph adjacency $\edgemat' = [\edgematsmall'_{ij}]$. Thus, $d'_{ii} = \sum_{j} w'_{ij}$ and $d'_{ij} = 0$ if $i\neq j$. Let $\lap' = \bD' - \edgemat'$ be the Laplacian matrix for the modified graph. \refeq{simpleobj} can be re-written in matrix format as:
\begin{equation}
\loss = \frac{1-\eta}{2} \norm{\bX-\bY}^2_F + \frac{\eta}{2} \bX^T\lap'\bX\label{eq:matrixobj1}
\end{equation}
We also consider a version of the loss function which uses the \textit{normalized} Laplacian $\symlap' = \bD'^{-1/2}\lap'\bD'^{-1/2}$ for regularization:
\begin{equation}
\tilde{\loss} = \frac{1-\eta}{2} \norm{\bX-\bY}^2_F + \frac{\eta}{2} \bX^T\symlap'\bX\label{eq:matrixobj}
\end{equation}
Using $\symlap'$ in place of $\lap'$ performs as well if not better in practice; and moreover provides certain theoretical guarantees (see \refpr{convergence}, and also \cite{von2008consistency}). Therefore, we will use \refeq{matrixobj} as the loss function for our higher-order label spreading and refer to it as $\loss_\holp$. 
A closed-form solution can now be obtained by differentiating $\loss_\holp$ with respect to $\bX$ and setting it to zero. Thus:
\begin{equation}
\bX = (1-\eta)\left(\identity - \eta (\identity-\symlap')\right)^{-1}\bY\label{eq:closedform}
\end{equation}
Thus, using \refeq{closedform}, we are able to compute the optimal solution to \holp, as long as the inverse of $\identity - \eta (\identity-\symlap')$ exists. 
Due to the use of normalized Laplacian regularization, the following holds: 
\begin{proposition}[Generalized Label Spreading]
	The proposed \hols algorithm reduces to traditional label spreading \cite{DBLP:conf/nips/ZhouBLWS03} for the base case of using only edge motifs, i.e., $\hons=\{K_2\}$.
\end{proposition}
This generalization grants \hols its name.
In practice, matrix inversion is computationally intensive and tends to be numerically unstable. Hence, we resort to an iterative approach to solve \refeq{closedform} by first initializing $\bX$ to an arbitrary value and then repeatedly applying the following update:
\begin{equation}
\bX \leftarrow \eta (\identity-\symlap')\bX + (1-\eta)\bY\label{eq:update}
\end{equation}
\refpr{convergence} describes the theoretical properties of this approach.
\begin{proposition}[Convergence Guarantee for \holp]\label{pr:convergence}
	The iterative update in \refeq{update} always converges to the unique fixed point given in \refeq{closedform} for any choice of initial $\bX$.
\end{proposition}

This can be proved using the theory of sparse linear systems~\cite{saad2003iterative}.

The overall algorithm of \holp is summarized in \refalg{hols}. First, for each motif $\kappa\in\hons$, construct its $\kappa$-participation matrix by enumerating all its occurrences. Note that the enumerated occurrences are processed one by one on the fly to update the participation matrix and discarded (no need for storage). Moreover, the enumeration for different motifs can be done in parallel. The participation matrices are combined into a single modified graph adjacency $\edgemat'$; applying the iterative updates from \refeq{update} finally results in labels for the unlabeled vertices. In practice, the iterative updates are applied until entries in $\bX$ do not change up to a precision $\epsilon$ or until a maximum number of iterations $T$ is reached.

\subsection{Complexity Analysis}
When only cliques are used as motifs $\hons$ for semi-supervised learning, the following space and time complexity bounds hold:
\begin{proposition}[Space Complexity of \holp]\label{pr:space}
	The space complexity of \holp\ for a graph with $N$ vertices, $M$ edges and $\nclass$ classes is $\oof{M + N\nclass}$ independent of motif size and number of motifs used, provided all motifs are cliques.
\end{proposition}

	
\begin{proposition}[Time Complexity of \holp]
	The time complexity of \holp\ over a graph with $M$ edges, $\nclass$ classes and a degeneracy (core number) of $k_{\max}$ using $\hons=\{K_2, \ldots, K_n\}$ is given by  $\oof{M\sum_{k=2}^{\maxcliquesize}k \left(\frac{k_{\max}}{2}\right)^{k-2}  }$ for the construction of $K_k$-participation matrices plus $\oof{M\nclass}$ per iterative update using \refeq{update}.
\end{proposition}

The proofs rely on the sparsity structure of the modified adjacency $\edgemat'$ and also Theorem 5.7 of \cite{DBLP:conf/www/DanischBS18}. 
Despite the exponential complexity in $k$, we are able to enumerate cliques quickly using the sequential \textsc{kClist} algorithm \cite{DBLP:conf/www/DanischBS18}. For example, our largest \pokec dataset has 21M edges, 32M triangles, 43M 4-cliques and 53M 5-cliques; and the enumeration of each took at most 20 seconds on a stock laptop. Thus, \hols remains fast and scalable when reasonably small cliques are used. Further, as we show in experiments, using triangles ($K_3$) in addition to edges typically suffices to achieve the best classification performance across a wide range of datasets.

\section{Experiments}
\label{sec:experiments}
We empirically evaluate the proposed algorithm on the four real-world network datasets described in \refsec{datasets}.

\subsection{Experimental Setup}
We implement \hols in MATLAB and run the experiments on MacOS with 2.7 GHz Intel Core i5 processor and 16 GB main memory.

\textbf{Baselines.} We compare \hols to the following baselines: \textit{(1) Label Propagation (\lp)} \cite{DBLP:conf/icml/ZhuGL03} which uses Laplacian regularization. \textit{(2) Label Spreading (\ls)} \cite{DBLP:conf/nips/ZhouBLWS03} which uses normalized Laplacian regularization. \textit{(3) \nvsvm} which generates unsupervised vertex embeddings using \nodetovec \cite{node2vec} and learns decision boundaries in the embedding space using one-vs-rest transductive SVMs \cite{tsvm}. \textit{(4) Graph Convolutional Network (\gcn)} \cite{iclr17gcn} which is an end-to-end semi-supervised learner using neural networks. We implement \lp and \ls in MATLAB, and use open-sourced code for the rest.

\textbf{Parameters.} By default, we use a weight of $\eta=0.5$ for supervised loss and $\hons=\{K_2,K_3\}$ motifs (edges and triangles) for \hols. The importance weight for triangles $\alpha_{K_3}$ is tuned in $\{0.1, 0.2, \ldots, 0.9\}$ for each dataset and results are reported on the best performing value. We use $\eta=0.5$ for \ls as well. \lp, \ls and \hols are run until labels converge to a precision of $\epsilon$ or until $T$ iterations are completed, whichever occurs sooner. We set $\epsilon=10^{-6}$ and $T=500$. We use the default hyperparameters for \gcn, \nodetovec and \tsvm. We supply 100, 20, 100 and 1000 labels for \euemail, \polblogs, \cora and \pokec datasets, where the vertices to label are chosen by stratified sampling based on class. These correspond to label fractions of 5\%, 1.6\%, 0.4\% and 0.06\% and on an average, 1, 10, 10 and 100 labeled vertices per class respectively.

\textbf{Evaluation metrics.} We quantify the success of semi-supervised learning using \textit{accuracy}, which is the fraction of unlabeled vertices which are correctly classified. We also note down the end-to-end \textit{running time} for all computation including any pre-processing such as clique enumeration, but excluding I/O operations.

\newcommand{\first}[1]{\textbf{\underline{#1}}}
\newcommand{\second}[1]{\underline{#1}}
\begin{table*}[!t]
	\centering
	\caption{Accuracy and Running Time (averaged over five runs):
		In each column, the best value is bold and underlined, and the second best is underlined.
		Asterisk ($^\ast$) denotes statistically significant differences ($p<0.05$) compared to the second best.
		\label{tab:accuracy}}
	\vspace{-2mm}
	\begin{tabular}{c|cccc|cccc}
		\toprule
		\multirow{2}{*}{\backslashbox{\textbf{Method}}{\textbf{Metric}}} & \multicolumn{4}{c|}{\textbf{Accuracy}} & \multicolumn{4}{c}{\textbf{Running time (seconds)}}\\
		& \textbf{\euemail} & \textbf{\polblogs} & \textbf{\cora} & \textbf{\pokec} & \textbf{\euemail} & \textbf{\polblogs} & \textbf{\cora} & \textbf{\pokec} \\
		\midrule
		Label Propagation (\lp) \cite{DBLP:conf/icml/ZhuGL03} & 0.2905 & 0.5814 & 0.2765 & 0.1994 & 0.11 & \second{0.070} & 2.1 & 1320 \\
		Label Spreading (\ls) \cite{DBLP:conf/nips/ZhouBLWS03} & 0.5228 & 0.9361 & \second{0.4921} & \second{0.5514} & \first{0.040}$^\ast$ & \first{0.036}$^\ast$ & \first{0.21}$^\ast$ & \first{93}$^\ast$\\
		\nvsvm \cite{node2vec,tsvm} & 0.4563 & \first{0.9481} & 0.4233 & T.L.E. & 46 & 29 & 3060 & >1 day\\
		Graph Convolution Networks (\gcn) \cite{iclr17gcn} & \second{0.5251} & 0.9470 & 0.4673 & 0.5290 & 1.8 & 1.3 & 6.4 & 2880\\
		\midrule
		\hols (proposed) & \first{0.5473}$^\ast$ & \second{0.9476} & \first{0.4953}$^\ast$ & \first{0.5593}$^\ast$ & \second{0.089} & 0.083 & \second{0.41} & \second{117}\\
		\bottomrule
	\end{tabular}
\end{table*}

\subsection{Results}
We present our experimental results. All reported values are averaged over five runs, each run differing in the set of labeled vertices.

\textbf{Accuracy.}
Accuracies of all methods are summarized in \reftab{accuracy}(left).
The values for \nvsvm on \pokec dataset are missing as the method did not terminate within 24 hours (`T.L.E.').

First, we observe in \reftab{accuracy}(left) that \hols consistently leads to statistically significant improvements over \ls, showing that using higher-order structures for label spreading helps.
In addition, \hols outperforms \textit{all} baselines in three out of four datasets.
The improvements over the best baseline are statistically significant $(p<0.05)$ according to a two-sided micro-sign test \cite{DBLP:conf/sigir/YangL99} in at least three out of five runs.
Importantly, for the smaller datasets (\euemail and \polblogs), while \gcn outperforms \ls, \gcn loses to \hols when triangles are used.
\nvsvm performs slightly better than \hols on \polblogs, however, the increase over \hols is not statistically significant.
For the larger datasets with $<0.5\%$ labeled vertices, \hols performs the best and \ls follows closely.

\textbf{Running Time.}
The running time of \hols and all the baselines is summarized in \reftab{accuracy}(right).
Notably, we see that \hols\ runs in less than 2 minutes for graphs with over 21 million edges (the \pokec graph), demonstrating its real-world practical scalability.

We observe that \ls is the fastest of all methods and \hols comes a close second for three out of four datasets.
The small difference in running time predominantly stems from the construction of triangle participation matrix.
Furthermore, \hols is over $15\times$ faster than the recent \gcn and \nvsvm baselines, \textit{for comparable and often better values of accuracy}.

\renewcommand{\mywidth}{0.45\columnwidth}
\begin{figure}
	\centering
	\resizebox{\columnwidth}{!}{
			\includegraphics[width=\mywidth]{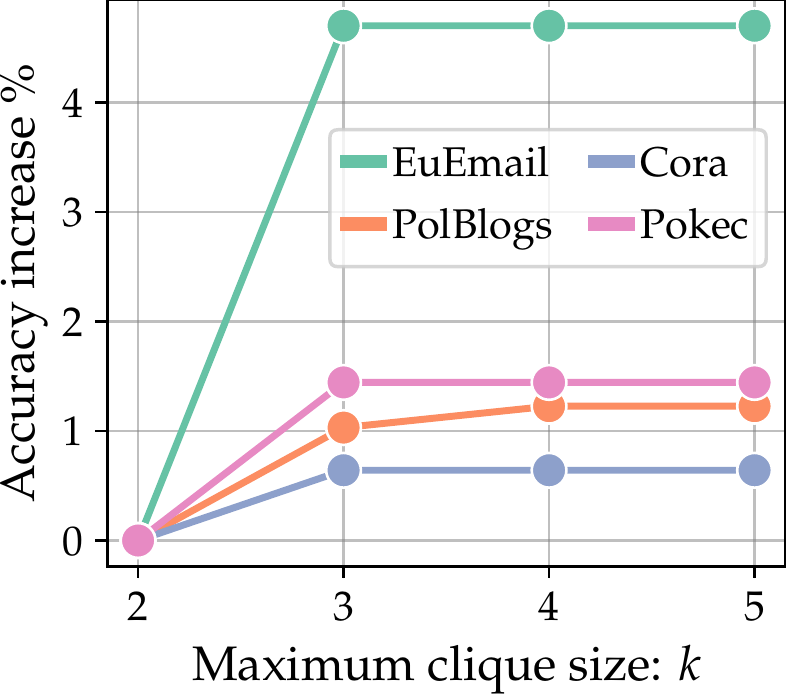} 
			\includegraphics[width=\mywidth]{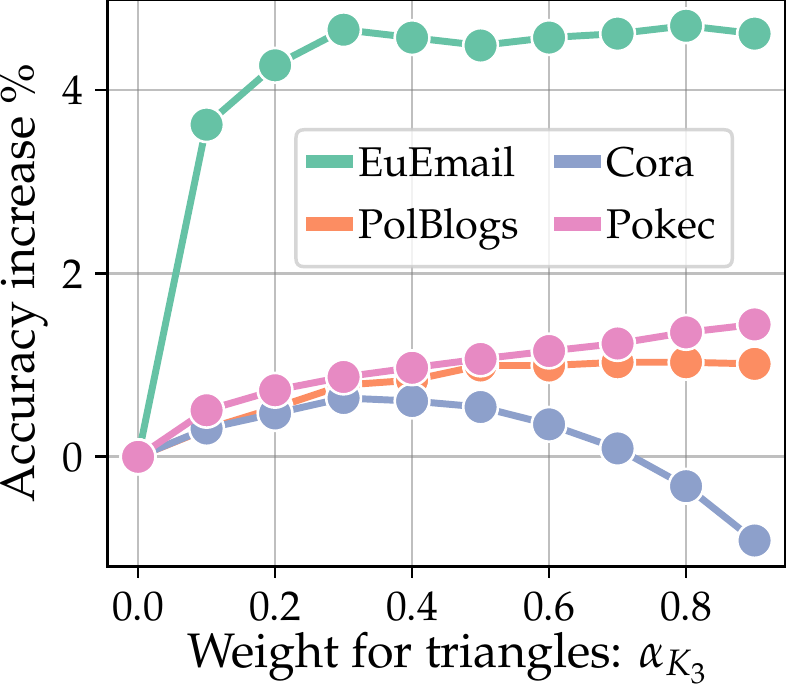} 
	}
	\caption{Variation of accuracy with maximum clique size $k$ (left), and importance weight $\alpha_{K_3}$ to triangles (right).
		\label{fig:parameters}}
\end{figure}

\renewcommand{\mywidth}{}
\begin{figure}
	\centering
		\includegraphics[width=0.4\linewidth]{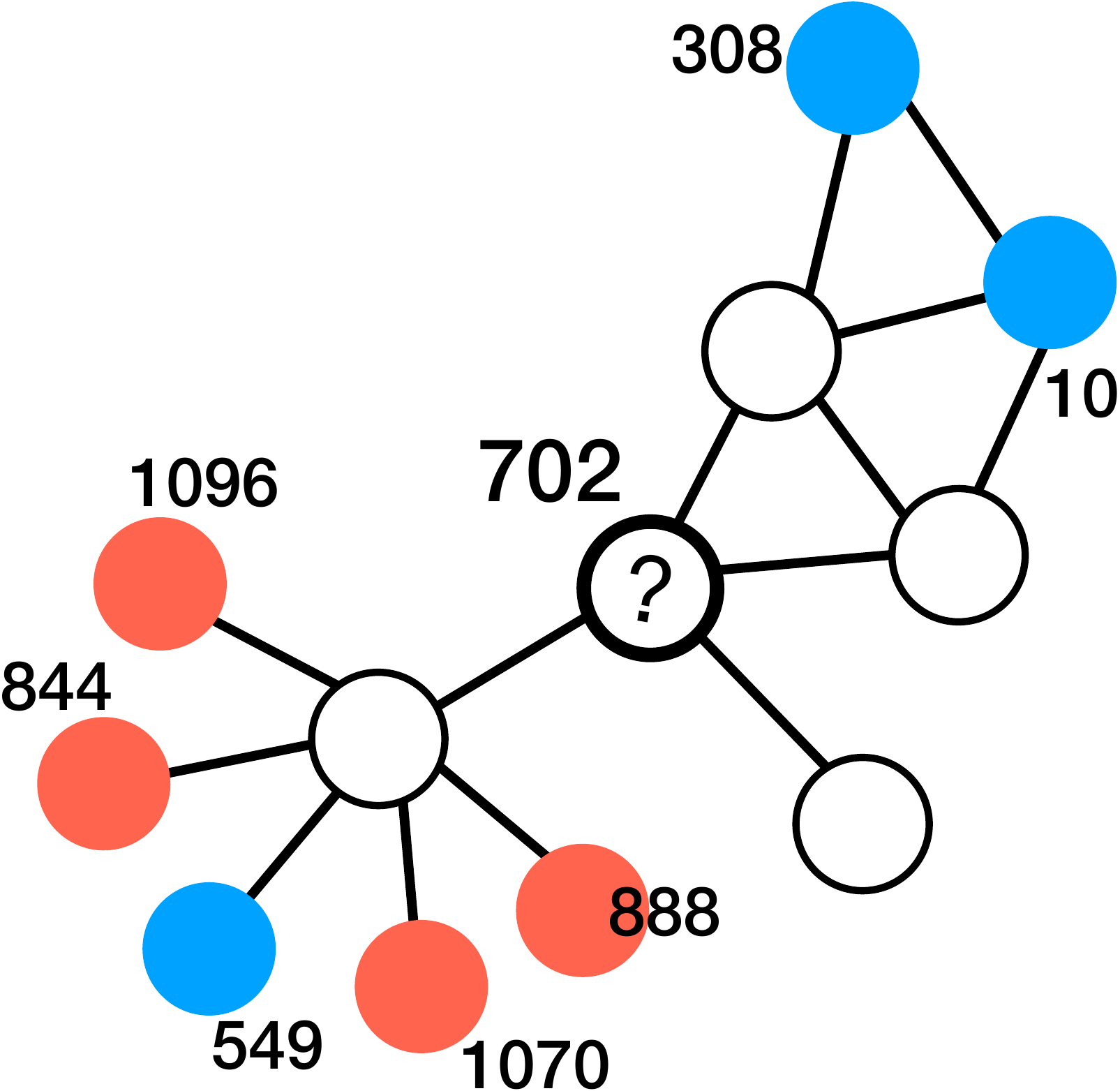} 
		\includegraphics[width=0.44\linewidth]{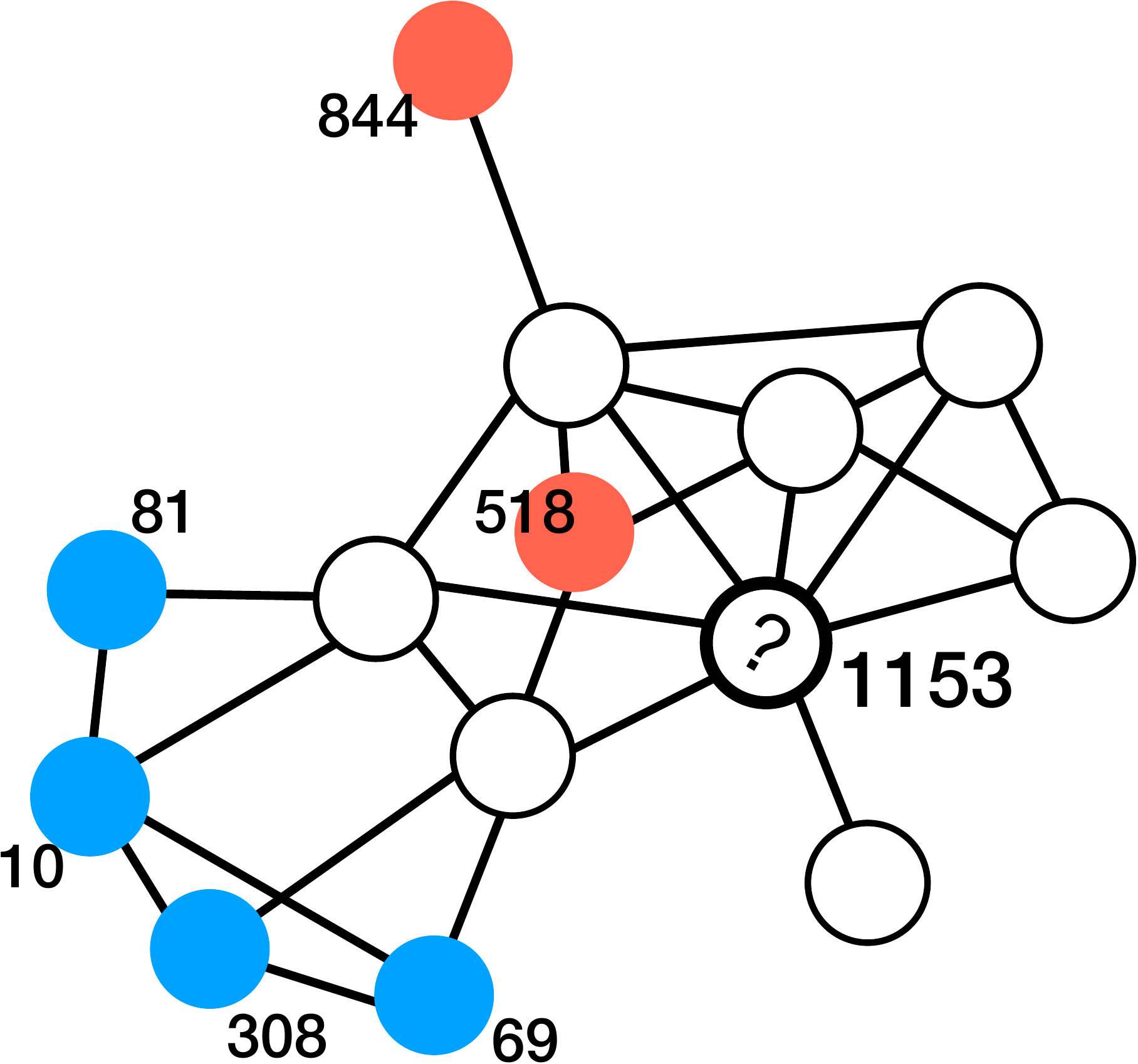} 
	\caption{Case studies from \polblogs dataset showing extended ego-networks of vertices v702 (left) and v1153 (right) which are both incorrectly classified by \ls but correctly classified when triangles are taken into account using \hols.\label{fig:casestudies}}
	\vspace{-2mm}
\end{figure}

\textbf{Accuracy vs. Maximum Clique Size.}
Fixing the motif set as $\hons=\{K_2, K_3, \ldots, K_{k}\}$, we vary $k=2, 3, 4, 5$ to study the marginal benefit of including higher-order cliques in graph SSL. The motif weights are tuned in $\alpha_{K_j} \in \{0, 0.1, \ldots, 0.9\}$, ensuring that edges are given a weight $\alpha_{K_2}\geq0.1$ for a connected graph, and further, all motif weights sum to 1. The best performing motif weights were used to generate \reffig{parameters}{(a)}, which plots the relative improvement in accuracy over \ls that uses edges only. We note that label spreading via higher-order structures strictly outperforms label spreading via edges. The gain is the most when using 3-cliques (triangles). Subsequent higher-order cliques did not lead to additional performance gain in most cases, presumably because their occurrences tend to be concentrated around a few high-degree vertices.

\textbf{Accuracy vs. Importance Weight To Triangles.}
Fixing the motif set to $\hons=\{K_2, K_3\}$, we vary the importance weight $\alpha_{K_3}$ to triangles in $\{0, 0.1, \ldots, 0.9\}$ to understand its effect on accuracy.
\reffig{parameters}{(b)} shows that the accuracy gain of \hols over \ls increases with an increase in triangle weight for most graphs.
The only exception is \cora, where the accuracy gain grows until $\alpha_{K_3}=0.4$ before decreasing and eventually turning negative.
Overall, triangles consistently help over a large range of motif weights.

\textbf{Case Studies.}
In \reffig{casestudies}, we look at real examples from the \polblogs\ dataset to dig deeper into when \hols improves over \ls.
Question mark denotes the central vertices v702 and v1153 of interest with ground truth labels `blue' and `red' respectively.
The direct neighbors of the both vertices are unlabeled and a few second hop neighbors are labeled with one of two labels: `blue' or `red'.

In both cases, both \ls and \hols classify the unlabeled 1-hop neighbors correctly. However, \ls, relying only on the edge-level information (roughly the ratio of blue to red labeled 2-hop neighbors in this case), incorrectly labels both v702 and v1153. The proposed \hols, on the other hand, accurately recognizes that v702 (v1153) is more tightly connected with its blue (red) neighbors via the higher-order triangle structures and thus leads to the correct classification. 



\section{Conclusion}
\label{sec:conclusion}
In this paper, we demonstrated that label homogeneity--the tendency of vertices participating in a higher-order structure to share the same label--is a prevalent characteristic in real-world graphs. We created an algorithm to exploit the signal present in higher-order structures for more accurate semi-supervised learning over graphs. Experiments on real-world data showed that using triangles along with edges for label spreading leads to statistically significant accuracy gains compared to the use of edges alone.

This work opens the avenue for several exciting future research directions. First, we need principled measures quantifying label homogeneity to aid comparison across diverse graphs and higher-order structures. Next, having seen the improvements in real-world graphs, it becomes fundamental to understand the benefits in random graph models. Finally, it is crucial to develop algorithms exploiting higher-order structures which can cater to the increasingly heterogeneous and dynamic nature of real-world graphs at scale.

\vspace{2mm}
\noindent{\small \textbf{Acknowledgments.} This material is based upon work supported by the National Science Foundation under Grants No. CNS-1314632 and IIS-1408924, by Adobe Inc., and by Pacific Northwest National Laboratory. Any opinions, findings, and conclusions or recommendations expressed in this material are those of the author(s) and do not necessarily reflect the views of the National Science Foundation, or other funding parties. The U.S. Government is authorized to reproduce and distribute reprints for Government purposes notwithstanding any copyright notation here on.}

\clearpage
\balance
\bibliographystyle{ACM-Reference-Format}
\bibliography{BIB/ref,BIB/data,BIB/ssl,BIB/clique}

\end{document}